# Intrinsic gain modulation and adaptive neural coding

Sungho Hong[1,2], Brian Nils Lundstrom[1] and Adrienne L. Fairhall[1]

[1]Physiology and Biophysics Department

University of Washington

Seattle, WA 98195-7290

[2]Current address: Computational Neuroscience Unit,

Okinawa Institute of Science and Technology

7542 Onna, Onna-son, Okinawa 904-0411

Japan

## Abstract

In many cases, the computation of a neural system can be reduced to a receptive field, or a set of linear filters, and a thresholding function, or gain curve, which determines the firing probability; this is known as a linear/nonlinear model. In some forms of sensory adaptation, these linear filters and gain curve adjust very rapidly to changes in the variance of a randomly varying driving input. An apparently similar but previously unrelated issue is the observation of gain control by background noise in cortical neurons: the slope of the firing rate vs current (*f-I*) curve changes with the variance of background random input. Here, we show a direct correspondence between these two observations by relating variance-dependent changes in the gain of *f-I* curves to characteristics of the changing empirical linear/nonlinear model obtained by sampling. In the case that the underlying system is fixed, we derive relationships relating the change of the gain with respect to both mean and variance with the receptive fields derived from reverse correlation on a white noise stimulus. Using two conductance-based model neurons that display distinct gain modulation properties through a simple change in parameters, we show that coding properties of both these models quantitatively satisfy the predicted relationships. Our results describe how both variance-dependent gain modulation and adaptive neural computation result from intrinsic nonlinearity.

## Author summary

A neuron receives an input whose statistical properties may change in time, and transforms it into an output. Many neurons are known to achieve a wide dynamic range by adaptively changing their computational function according to the input statistics. These adaptive changes can be very rapid and it has been suggested that a component of this adaptation could be purely input-driven: even a fixed neural system can show apparent adaptive behavior since inputs with different statistics interact with the nonlinearity of the system in different ways. In this paper, we show how a single neuron's intrinsic computational function can dictate such input-driven changes in its response to varying input statistics. We show how two different characterizations of neural function—in terms of mean firing rate and in terms of generating precise spike timing—are related. We then apply our results to two biophysically defined model neurons, which have significantly different response patterns to inputs with various statistics. We show that their behaviors can be well explained by our model of intrinsic adaptation. Contrary to the picture that neurons carry out a stereotyped computation on their inputs, our results show that even in the simplest cases they have simple yet effective mechanisms by which they can adapt to their input. Adaptation to stimulus statistics, therefore, is built into the most basic single neuron computations.

## Introduction

An *f-I* curve, defined as the mean firing rate in response to a stationary mean current input, is one of the simplest ways to characterize how a neuron transforms a stimulus into a spike train output as a function of the magnitude of a single stimulus parameter. Recently, the dependence of *f-I* curves on other input statistics such as the variance has been examined: the slope of the *f-I* curve, or gain, is modulated in diverse ways in response to different intensities of added noise [1-4]. This enables multiplicative control of the neuronal gain by the level of background synaptic activity [1]: changing the level of the background synaptic activity is equivalent to changing the variance of the noisy balanced excitatory and inhibitory input current to the soma, which modulates the gain of the *f-I* curve. It has been demonstrated that such somatic gain modulation, combined with saturation in the dendrites, can lead to multiplicative gain control in a single neuron by background inputs [5]. From a computational perspective, the sensitivity of the firing rate to mean or variance can be thought of as distinguishing the neuron's function as either an integrator (greater sensitivity to the mean) or a differentiator/coincidence detector (greater sensitivity to fluctuations, as quantified by the variance) [3,6,7].

An alternative method of characterizing a neuron's input-to-output transformation is through a linear/nonlinear (LN) cascade model [8,9]. These models comprise a set of linear filters or receptive field that selects particular features from the input; the filter output is transformed by a nonlinear threshold stage into a time-varying firing rate. Spike-triggered covariance analysis [10,11] reconstructs a model with multiple features from a neuron's input/output data. It has been widely employed to characterize both neural systems [12-15] and single neurons or neuron models subject to current or conductance inputs [16-19].

Generally, results of reverse correlation analysis may depend on the statistics of the stimulus used to sample the model [15,19-25]. While some of the dependence on stimulus statistics in the response of a neuron or neural system may reflect underlying plasticity, in some cases, the rapid timescale of the changes suggests the action of intrinsic nonlinearities in systems with *fixed* parameters [16,19,25-29], which changes the *effective* computation of a neuron.

Our goal here is to unify the *f-I* curve description of variance-dependent adaptive computation with that given by the LN model: we present analytical results showing that the variance-dependent modulation of the firing rate is closely related to adaptive changes in the *recovered* LN model if a fixed underlying model is assumed. When the model relies only on a single feature, we find that such a system can show only a single type of gain modulation, which accompanies an interesting asymptotic scaling behavior. With multiple features, the model can show more diverse adaptive behaviors, exemplified by two conductance-based models that we will study.

## Results

### Diverse variance-dependent gain modulations without spike rate adaptation

Recently, Higgs et al. [3] and Arsiero et al. [4] identified different forms of variance-dependent change in the *f-I* curves of various neuron types in avian brainstem and in cortex. Depending on the type, neurons can have either increasing or decreasing gain in the *f-I* curve with increasing variance. These papers linked the phenomenon to mechanisms underlying spike rate adaptation, such as slow afterhyperpolarization (sAHP) currents and slow sodium channel inactivation. We recently showed [7] that a standard Hodgkin-Huxley (HH) neuron model, lacking spike rate adaptation, can show two different types of variance-dependent gain modulation simply by tuning the maximal conductance parameters of the model. These differences in gain modulation correspond to two different regimes in the space of conductance parameters. In one regime, which includes the standard parameters, a neuron periodically fires to a sufficiently large constant input current. In the other regime, a neuron never fires to a constant input regardless of its magnitude, but responds only to rapid fluctuations. This rarely discussed property has been termed *class 3 excitability* [30,31]. Higgs et al. [3] proposed that the type of gain modulation classifies the neuron as an integrator or differentiator.

Here, we examine two models that show these different forms of variance-dependent gain modulation without spike rate adaptation, and study the resulting LN models sampled with different stimulus statistics. We show that these *fixed* models generate variance-dependent gain modulation, and that this gain modulation is well predicted by aspects of the LN models derived from white noise stimulation. The two models are both based on the Hodgkin-Huxley (HH) [32] active currents; one model is the standard HH model, and the other (HHLS) has lower $Na^+$ and higher $K^+$ conductances. The HHLS model is a class 3 neuron and responds only to a rapidly changing input. For this reason, the HHLS model can be thought of as behaving more like a differentiator than an integrator [3,7].

Fig. 1 shows the different gain modulation behaviors of the HH and HHLS conductance-based models. For the HH model, Fig. 1A, the *f-I* curves in the presence of noise are similar to the noiseless case except that they are increasingly smoothed at the threshold. In contrast, Fig. 1C shows that the *f-I* curves of the HHLS model never converge toward each other as the noise level increases. This case resembles that of layer 5 pyramidal neurons in rat medial prefrontal cortex [4], as well as nucleus laminaris (NL) neurons in the chick auditory brainstem and some pyramidal neurons in layer 2/3 of rat neocortex [3]. While for these layer 2/3 neurons, there is evidence that this change in *f-I* curve slope may be related to the sAHP current [3], at steady state this effect can be obtained in general by tuning the maximal conductances without introducing any mechanism for spike rate adaptation [7].

**Gain modulation and adaptation of fixed models**

For a system described by an LN model with a single feature, we derive an equation relating the slopes of the firing rate with respect to stimulus mean and variance. We then consider gain modulation in a system with multiple relevant features and derive a series of equations relating gain change to properties of the spike-triggered average and spike-triggered covariance. Throughout, we assume that the underlying system is fixed, and that its parameter settings do not depend on stimulus statistics. For example, if the model has a single exponential filter with a time constant $\tau$, we assume that $\tau$ does not change with the stimulus mean ($I_0$) or variance ($\sigma^2$). However, this does not mean that the model shows a single response pattern regardless of the statistical structure of stimuli. The sampled LN description of a nonlinear system with fixed parameters—even when the underlying model is an LN model [25]—can show interaction with the input statistics, leading to different LN model descriptions for different input parameters [19,25,27-29]. We refer to this as *intrinsic adaptation*.

**One-dimensional model**

An LN model is composed of its relevant features $\{\varepsilon_\mu(t)\}, \quad (\mu = 1, 2, \ldots, n)$, which act as linear filters on an incoming stimulus, and a probability to spike given the filtered stimulus, $P(\text{spike}|\text{filtered stimulus})$. For a Gaussian white noise stimulus with mean $I_0$ and variance $\sigma^2$, the firing rate is

$$f(I_0,\sigma^2) = \int d\mathbf{x}\, P(\text{spike} \mid I_0\bar{\boldsymbol{\varepsilon}} + \mathbf{x})\, p(\mathbf{x}), \tag{1}$$

where $\bar{\boldsymbol{\varepsilon}} = \int_0^\infty \boldsymbol{\varepsilon}(\tau)d\tau$ is the time-integrated filter and $\mathbf{x}$ is the mean-subtracted noise stimulus filtered by the $n$ relevant features. $p(\mathbf{x})$ is an $n$-dimensional Gaussian distribution with variance $\sigma^2$. We refer to the Materials and Methods for a more detailed account of the model.

For a one-dimensional model $n=1$, Eq. (1) can be rewritten with change of variables

$$f(I_0,\sigma^2) = \int_{-\infty}^{\infty} dx\, P(\text{spike} \mid x)\, p(x - I_0\bar{\varepsilon}). \tag{2}$$

Since $p(x)$ is Gaussian, it is also the kernel or Green's function of a diffusion equation in terms of $(x,\sigma^2)$, and therefore so is $p(x-I_0\bar{\varepsilon})$ in terms of $(I_0,\sigma^2)$. In other words, we have

$$\left(\frac{\partial}{\partial\sigma^2} - \frac{1}{2}\frac{\partial^2}{\partial x^2}\right) p(x - I_0\bar{\varepsilon})$$
$$= \left(\frac{\partial}{\partial\sigma^2} - \frac{1}{2\bar{\varepsilon}^2}\frac{\partial^2}{\partial {I_0}^2}\right) p(x - I_0\bar{\varepsilon}) = 0.$$

Now operating with $\left(\frac{\partial}{\partial\sigma^2} - \frac{1}{2\bar{\varepsilon}^2}\frac{\partial^2}{\partial {I_0}^2}\right)$ on both sides of the equation, $p(x-I_0\bar{\varepsilon})$ is the only term on the left hand side of Eq. (2) that depends on $(I_0,\sigma^2)$ and therefore the right hand side of Eq. (2) vanishes. Thus one finds

$$2\bar{\varepsilon}^2 \frac{\partial f}{\partial\sigma^2} = \frac{\partial^2 f}{\partial {I_0}^2}. \tag{3}$$

The boundary condition is given by evaluating Eq. (2) as $\sigma^2 \to 0$; here the Gaussian distribution becomes a delta function,

$$\lim_{\sigma^2\to 0} p(x - I_0\bar{\varepsilon}) = \delta(x - I_0\bar{\varepsilon}),$$

and the boundary condition is given by the zero-noise *f-I* curve. Thus, when a model depends only on a single feature, $\varepsilon(t)$, the *f-I* curve with a noisy input is given by a simple diffusion-like equation, Eq. (3), with a single parameter, the time integrated filter, $\bar{\varepsilon} = \int_0^\infty \varepsilon(\tau)d\tau$, determining the diffusion constant $1/2\bar{\varepsilon}^2$.

Eq. (3) states that the variance-dependent change in the firing rate is simply determined by the curvature of the *f-I* curve. Thus, a one-dimensional system displays only a single type of noise-induced gain modulation: as in diffusion, an *f-I* curve is gradually smoothed and flattened as the variance increases. Given a boundary condition, such as an *f-I* curve for a particular variance, the family of *f-I* relations can be reconstructed up to a scale factor by solving Eq. (3). For example, one can predict how the neuron would respond to a noise stimulus based on its output in the absence of noise. Note that the solution of Eq. (3) generalizes a classical result [33] based on a binary nonlinearity to a simple closed form which applies to any type of nonlinearity.

Figs. 2A and B show a solution of Eq. (3). While this one-dimensional model is based on the simplest and most general assumptions, it provides insights into the structure of variance-dependent gain modulation. The boundary condition is an *f-I* curve with no noise, $f = \sqrt{I + 0.1}$ for $I > 0$ and $f = 0$ for $I \leq 0$, which imitates the general behavior of many dynamical neuron models around rheobase [34-36]. Compared with the HH conductance-based model, Eq. (3) captures qualitative characteristics of the HH *f-I* curve despite differences due to the increased complexity of the HH model over a 1D LN model: in Fig. 2A and B, there is a positive curvature (second derivative of firing rate with respect to current) of the *f-I* curve below rheobase related to the increase of the firing rate with increasing variance. In contrast, the behavior of the HHLS model cannot be described by Eq. (3). Even though the *f-I* curves in Fig. 1C mostly have negative curvature, the firing rate keeps increasing with variance, implying that the HHLS model cannot be described by a one-dimensional LN model.

We also compared Eq. (3) with the *f-I* curves from two commonly used simple neuron models, the leaky integrate-and-fire (LIF) model (Fig. 2C), and a similar model with minimal nonlinearity, the quadratic integrate-and-fire (QIF) model [37,38] (Fig. 2D). The *f-I* curves of the two models are similar but have subtle differences: in the LIF model, firing rate never decreases with noise, even though parameters were chosen to induce a large negative curvature, as shown analytically in Supporting Information. The QIF model behavior is much more similar to the 1D LN model, marked by a slight decrease in firing rate at large $I_0$. From this perspective, the QIF is a *simpler* model in terms of the LN description despite the dynamical nonlinearity.

It is interesting to note that for one-dimensional models, the gain modulation given by Eq. (3) depends only on the boundary condition, which implicitly describes how an input with a given mean samples the nonlinearity, but not explicitly on the details of filters or nonlinearity. An ideal differentiator, where firing rate is independent of the stimulus mean, is realized only when the filter has zero integral, $\bar{\varepsilon} = 0$. This is also the criterion that would be satisfied if the filter itself were ideally differentiating. We will return to the relationship between the LN model functional description and that of the *f-I* curves in the Discussion.

**Multi-dimensional models**

Here we examine gain modulation in the case of a system with multiple relevant features. In this case, one cannot derive a single simple equation such as Eq. (3). Instead, we derive relationships between the characteristics of $f(I_0, \sigma)$ curves and quantities calculated using white noise analysis.

Fixed multi-dimensional models can display far more complex response patterns to different stimulus statistics than one-dimensional models, because linear components in the model can now interact nonlinearly [29]. For example, in white noise analysis, as the stimulus variance increases, the distribution of the filtered stimuli also expands and probes different regions of the nonlinear threshold structure of the model. This induces a variance-dependent rotation among the filters recovered through sampling by white noise analysis, and the corresponding changes in the spike-triggered average, spike-triggered covariance, and the sampled nonlinearity [19].

Here, we relate parameters of the changing spike-triggered average and spike-triggered covariance description to the form of the *f-I* curves. The relationships are derived by taking derivatives of each side of Eq. (1) with respect to $I_0$ and $\sigma^2$ (see Materials and Methods). The first order in $I_0$ establishes the relationship between the STA and the gain of the *f-I* curve with respect to the mean:

$$\frac{\partial \log f}{\partial I_0} = \frac{1}{\sigma^2}\overline{\mathrm{STA}}, \quad \overline{\mathrm{STA}} = \int_0^\infty d\tau\, \mathrm{STA}(\tau). \tag{4}$$

The second order leads to a relationship between the second derivative of the *f-I* curve and the covariance matrix:

$$\frac{\partial^2 \log f}{\partial {I_0}^2} = \frac{1}{\sigma^4}\overline{\Delta C}, \quad \overline{\Delta C} = \int d\tau d\tau' \Delta C(\tau,\tau'). \tag{5}$$

The gain with respect to the variance is

$$\frac{\partial \log f}{\partial \sigma^2} = \frac{1}{2\sigma^4}\left(\mathrm{Tr}\Delta C + \left\|\mathrm{STA}\right\|^2\right), \tag{6}$$

where

$$\mathrm{Tr}\Delta C = \int d\tau\, \Delta C(\tau,\tau), \quad \left\|\mathrm{STA}\right\|^2 = \int d\tau\, \mathrm{STA}(\tau)^2.$$

Eqs. (4)–(6)show how the nonlinear gain of an *f-I* curve with respect to input mean and variance is related to intrinsic adaptation as observed through changes in the STA and STC. Note that Eqs. (4) -(6)apply to one-dimensional LN models as well. In that case, the STA has the same shape as the feature in the model, and only its magnitude varies according to the overlap integral, Eq. (1), between the nonlinearity of the model and the prior stimulus. This is the same for the STC, and thus Eqs. (4)-(6)are not independent. This leads to a single form of variance gain modulation, given by Eq. (3). However, in a multi-dimensional model, changing the stimulus mean shifts the nonlinearity in a single direction, $\overline{\mathrm{STA}}$, while increasing the variance expands the prior in every direction in the stimulus space. Therefore, the overlap integral can show more diverse behaviors.

**Conductance based models**

We now examine whether the gain modulation behaviors we have described can be captured by a multi-dimensional LN model. We tested this by computing *f-I* curves, spike-triggered averages and the spike-triggered covariance matrices for the noise-driven HH and HHLS models for a range of input statistics. Figures 3A, B, and C show the result of fitting simulation data from the HH (left) and HHLS (right) model to Eqs. (4), (5), and(6) respectively. The linear relationships are quite clear in Fig. 3A and C which show the gains with respect to mean and variance. Fig. 3B involves the curvature of *f-I* curves, which is more difficult to calculate accurately, and shows larger errors. In every case, goodness of fit is $p < 1.3 \cdot 10^{-6}$ and $p < 5.8 \cdot 10^{-6}$ for the HH and HHLS where the upper bounds of p-values are given by the case of Eq. (5), corresponding to Fig. 3B. These results show that intrinsic adaptation of the LN model predicts the form of noise-induced gain modulation for these models.

**Gain rescaling of one-dimensional models**

Here we discuss a consequence of intrinsic adaptation for neuronal encoding of mean and variance information for a one-dimensional model. In this case, Eq. (3) completely specifies intrinsic adaptation, and therefore we will focus on this case.

Our first observation is that Eq. (3) is invariant under the simultaneous rescaling of the mean and standard deviation, $I_0 \to \alpha I_0, \sigma \to \alpha\sigma$, where $\alpha$ is an arbitrary positive number. This invariance is preserved if the solution is also a function of only a dimensionless variable $I_0 / \sigma$, which would represent a signal-to-noise ratio if we describe the neuron's input/output function in terms of an *f-I* curve at a fixed noise level $\sigma$. Note that this situation is analogous to the Weber-Fechner [39,40] and Fitts' law [41], which states that perception tends to depend on only dimensionless variables that are invariant under scaling of the absolute magnitude of stimulus [42]. However, the invariance of Eq. (3) under the scaling of a stimulus does not necessarily lead to the invariance of a firing rate solution. By rewriting Eq. (2) in terms of the "rescaled" variables, $y = x / \sigma$ and $\mu = I_0 / \sigma$, we get

$$f(\mu,\sigma^2) = \frac{1}{\sqrt{2\pi}} \int dy\, e^{-(y-\mu\tilde{\sigma})^2/2} f_0\left(\frac{y\sigma}{\tilde{\sigma}}\right). \tag{7}$$

where $f_0(I) = P(\text{spike} \mid I\tilde{o})$ is an *f-I* curve with no noise. Thus, the scaling of $f(I_0, \sigma^2)$ with standard deviation depends on the boundary condition, $f_0(I)$, which in principle can be any arbitrary function.

Nevertheless, in practice, the *f-I* curves of many dynamical neurons are not completely arbitrary but can share a simple scaling property, at least asymptotically. For example, in the QIF and many other neuron models, the *f-I* curve with no noise asymptotically follows a power law $f_0 : \sqrt{I_0 - I_c}$ around the rheobase $I_c$ [34-36]. In general, if $f_0(I) \propto I^{\alpha}$ asymptotically in such a regime, from Eq. (7), the firing rate is asymptotically factorized into a $\sigma$ dependent and $\mu = I_0 / \sigma$ dependent part as

$$f(\mu, \sigma^2) \propto \sigma^{\alpha} F(\mu), \quad F(\mu) = \frac{1}{\sqrt{2\pi}} \int dy\, e^{-(y-\mu\tilde{o})^2/2} y^{\alpha} \,. \tag{8}$$

In other words, $I_0 / \sigma$ becomes an *intermediate asymptotic* of the *f-I* curves [43].

To test to what extent this scaling relationship holds in the models we have considered, we calculated the *rescaled relative gain* of the *f-I* curves, which we define as $(\sigma / f) \cdot \partial f / \partial I_0 = \sigma \cdot \partial \log f / \partial I_0$; the rescaled relative gain of Eq. (8) depends only on $\mu = I_0 / \sigma$, not on $\sigma$. Thus, if the rescaling strictly holds, this becomes a single-valued function of the signal-to-noise ratio, $I_0 / \sigma$, regardless of the noise level $\sigma$.

We find evidence for this form of variance rescaling in the QIF, LIF and HH models. Fig. 4 shows the rescaled gains evaluated from the simulated data. The QIF and HH case, Fig. 4B and D, match well with the solution of Eq. (3), Fig. 4A. In the LIF case, Fig. 4C, the relative gain shows deviations at low variance, but it approaches a variance-independent limit at large $\sigma$. We also present an analytic account in Supporting Information. On the other hand, in Fig. 4E, the HHLS model does not exhibit this form of asymptotic scaling at all. The role of the signal-to-noise ratio, $I_0 / \sigma$, in the HHLS model appears to be quite distinct from the other models. In summary, Eq. (3)

predicts that one-dimensional LN models will have the tendency to decrease gain with increasing noise level. However, if the *f-I* curve of a neuron is power-law-like, the resulting gain modulation will be such that the neuron's sensitivity to mean stimulus change at various noise levels is governed only by the signal-to-noise ratio.

## Discussion

In this paper, we have obtained analytical relationships between noise-dependent gain modulation of *f-I* curves and properties of the sampled linear/nonlinear model. We have shown that gain control arises as a simple consequence of the nonlinearity of the LN model, even with no changes in any underlying parameters.

For a system described by an LN model with only one relevant feature, a simple single-parameter diffusion relationship relates the *f-I* curves at different variances, where the role of the diffusion coefficient is taken by the integral of the STA. This form strictly limits the possible forms of gain modulation that may be manifested by such a system. The result qualitatively describes the variance dependent gain modulation of different neuron models such as the LIF, QIF, and standard HH neuron models. Models based on dynamical spike generation, such as QIF, showed better agreement with this result than the LIF model. The QIF model case is a good example of how a nonlinear dynamical system can be mapped onto an LN model description [19,44]. The QIF model has a single dynamical equation whose subthreshold dynamics are captured approximately by a linear kernel, which takes the role of the feature; one can then determine a threshold which acts as a binary decision boundary for spiking. Thus, it is reasonable that the QIF model and the one-dimensional LN model show a similar response pattern to a noisy input. When the system has multiple relevant features, we obtain equations relating the gain with respect to the input mean and the input variance to parameters of the STA and STC. We verified these results using Hodgkin-Huxley neurons displaying two different forms of noise-induced gain control.

Previous work has related different gain control behaviors to a neuron's function as an integrator or a differentiator [3,7]. From an LN model perspective, the neuron's function is defined by specific properties of the filter or filters $\boldsymbol{\varepsilon}(t)$. An integrating filter would consist of entirely positive weights; for leaky integrators these weights will decay at large negative times. A differentiating filter implements a local subtraction of the stimulus, and so should consist of a bimodal form where the positive weights approximately cancel the negative weights.

In general, characterizations of neural function by LN model and by *f-I* curves are quite distinct. The *f-I* approach we have discussed here describes the encoding of stationary statistical properties of the stimulus by time-averaged firing rate, while the LN model describes the encoding of specific input fluctuations by single spikes, generally under a particular choice of stimulus statistics. Indeed, the LN characterization can change with the driving stimulus distribution, even, in principle, from an integrator to a differentiator. Thus, a model may, for example, act as a differentiator on short timescales but as an integrator on longer timescales. For systems whose LN approximation varies with mean and variance, the neuron's effective computation changes with stimulus statistics, and so does the information that is represented. One might then ask how the system can decode the represented information. It has been proposed that statistics of the spike train might provide the information required to decode slower-varying stimulus parameters [22,45]. The possibility of distinguishing between responses to different stimulus statistics using the firing rate alone depends on the properties of the *f-I* curves.

The primary focus of this work is the restricted problem of single neurons responding to driving currents, where the integrated synaptic current in an *in vivo*-like condition is approximated to be a (filtered) Gaussian white noise [46-50]. However, our derivations can apply to arbitrary neural systems driven by white noise inputs, if *f-I* curves are interpreted as tuning functions with respect to the mean stimulus parameter. Given the generality of our results for neural systems, it would be interesting to test our results in cases where firing is driven by an external stimulus. A good candidate would be retinal ganglion cells, which are well-described by LN-type models [9,14,51-53], show adaptation to stimulus statistics on multiple timescales [23,54] and display a variety of dimensionalities in their feature space [14].

A limitation of the tests we have performed here is a restriction to the low firing rate regime where spike-triggered reverse correlation captures most of the dependence of firing probability on the stimulus. The effects of interspike interaction can be significant [16,17,55] and models with spike history feedback have been developed to account for this [44,51,56,57]. We have not investigated how spike history effects would impact our results.

Although evidence suggests that gain modulation by noise may be enhanced by slow afterhyperpolarization currents underlying spike frequency adaptation [3], these slow currents are not required to generate gain enhancement in simple neuron models [7,19,25-29]. While one may generate diverse types of noise-induced gain modulation only by modifying the mechanism of generating a spike independent of spike history [7], in realistic situations, slow adaptation currents are present and will affect neural responses over many timescales [58-60]. In principle, it is possible to extend our result to include these effects: *f-I* curves under conditions of spike frequency adaptation have been already discussed [61-63] and can be compared to LN models with spike history feedback. However, our goal here was to demonstrate the effects that can occur independent of slow adaptation currents and before such currents have acted to shift neuronal coding properties.

The suggestive form of our result for one-dimensional LN models led us to look for a representation of neuronal output that is invariant under change in the input noise level. Our motivation is based on a simple principle of dimensional analysis: the gains of the *f-I* curves with noise may be asymptotically described by a signal-to-noise ratio, a dimensionless variable depending only on the stimulus itself. We showed that this may occur if the *f-I* curve with no noise obeys asymptotic power-law properties. Such a property has been determined to arise both from the bifurcation patterns of spike generation [31,34,35] and due to spike rate adaptation [61]. This relationship implies that the gain of the firing rate as a function of the mean should scale inversely with the standard deviation. Scaling of the gain of the nonlinear decision function with the stimulus standard deviation has been observed to some degree in a number of neural systems [10,15,22-25,29,64-67]. Such scaling guarantees maximal transmission of information [10,22]. As we and others have proposed, a static model might suffice to explain this phenomenon [25,27], although in some cases slow adaptation currents are known to contribute [65,66].

In summary, we have presented theoretically derived relationships between the variance-dependent gain modulation of *f-I* curves and intrinsic adaptation in neural coding. In real neural systems, any type of gain modulation likely results from many different mechanisms, possibly involving long-time scale dynamics. Our results show that observed forms of gain modulation may be a result of a pre-existing static nonlinearity that reacts to changes in the stimulus statistics robustly and almost instantaneously.

## Materials and Methods

### Biophysical models

We used two single compartmental models with Hodgkin-Huxley (HH) active currents. The first one is an HH model with standard parameters while the second model (HHLS) has a lower $Na^+$ and higher $K^+$ maximal conductance. The voltage changes are described by [32]:

$$C\frac{dV}{dt} = -\bar{g}_L(V - E_L) - \bar{g}_{Na} m^3 h (V - E_{Na}) - \bar{g}_K n^4 (V - E_K) + I(t),$$

and the activation variables $m$, $n$ and $h$ behave according to

$$\tau_z(V)\frac{dz}{dt} = \bar{z}(V) - z, \qquad \tau_z = \frac{1}{\alpha_z + \beta_z}, \qquad \bar{z} = \frac{\alpha_z}{\alpha_z + \beta_z}, \qquad z = m, n, h.$$

where

$$\alpha_m = \frac{.1(V+40)}{1-\exp[-.1(V+40)]}, \qquad \beta_m = 4\exp[-.0556(V+65)],$$

$$\alpha_h = .07\exp[.05(V+65)], \qquad \beta_h = \frac{1}{1+\exp[-.1(V+35)]},$$

$$\alpha_n = \frac{.01(V+55)}{1-\exp[-.1(V+55)]}, \qquad \beta_n = .125\exp[-.0125(V+65)].$$

The voltage $V$ is in millivolts (mV).

For the HH model, the conductance parameters are $\bar{g}_K = 36\,\text{mS/cm}^2$ and $\bar{g}_{Na} = 120\,\text{mS/cm}^2$. The HHLS model has $\bar{g}_K = 41\,\text{mS/cm}^2$ and $\bar{g}_{Na} = 79\,\text{mS/cm}^2$. All other parameters are common to both models. The leak conductance is $\bar{g}_L = 0.3\,\text{mS/cm}^2$ and the membrane capacitance per area $C$ is $1\,\mu\text{F/cm}^2$. The reversal potentials are $E_L = -54.3\text{mV}$, $E_{Na} = 50\text{mV}$, and $E_K = -77\text{mV}$. The membrane area is $10^{-3}\,\text{cm}^2$, so that a current density of $1\,\mu\text{A/cm}^2$ corresponds to a current of 1nA.

All simulations of these models were done with the NEURON simulation environment [68]. Gaussian white noise currents with various means and variances are generated with an update rate of 5kHz ($dt = 0.2$ ms) and delivered into the model via current clamp. For the *f-I* curves, we simulated 4 minutes of input for each mean and variance pair. The whole procedure was repeated five times to estimate the variance of the *f-I* relationship, $\sigma_{\text{repeat}}$.

We ran another set of simulations for reverse correlation analysis and collected about 100,000 spikes for each stimulus condition. The means and variances of the Gaussian noisy stimuli were chosen such that the mean firing rate did not exceed 10Hz, and we selected eight means and seven variances for the HH model, and nine means and four variances for the HHLS model.

**Integrate-and-fire type models**

In addition to the conductance-based model, we investigated the behavior of two heuristic model neurons driven by a noisy current input. Each model consists of a single dynamical equation describing voltage fluctuations of the form

$$C\frac{dV}{dt} = L(V) + I(t).$$

The first model is a leaky integrate-and-fire (LIF) model [69,70], for which $L(V) = -g_L(V - E_L)$. We used the parameters $g_L = 2$, $E_L = 0$ and $C = 1$. Since this choice of $L(V)$ cannot generate a spike, we additionally imposed a spiking threshold $V_{th} = 1$ and reset voltage $V_{reset} = -3$.

The second is a quadratic integrate-and-fire (QIF) model [31,37,38], for which $L(V) = g_L(V - E_L)(V - V_{th}) / \Delta V$ where $\Delta V = V_{th} - E_L > 0$. We used $g_L = 0.5$, $E_L = 0$, $V_{th} = 0.1$, and $C = 1$. In this model, the voltage $V$ can increase without bound; such a trajectory is defined to be a spike if it crosses $V_{spike} = 5$. After spiking, the system is reset to $V_{reset} = 0$.

These two models are simulated using a fourth-order Runge-Kutta integration method with an integration time step of $dt = 0.01$. The input current $I(t)$ was Gaussian white noise, updated at each time step, with a range of means and variances. The *f-I* curves were obtained from 1000 sec of stimulation for each (mean,variance) condition. We then compared the *f-I* curves from these models with the relationship derived in the Results, Eq. (5). A numerical solution of the partial differential equation was obtained using a PDE solver in Mathematica (Wolfram Research Inc., Champaign, IL).

**Linear/nonlinear model**

We use the linear/nonlinear (LN) cascade model framework to describe a neuron's input/output relation. We will focus on the dependence of the firing rate of a fixed LN model on the mean and variance of a Gaussian white noise input.

We will take the driving input to be $I(t) = I_0 + \xi(t)$ where $I_0$ is the mean and $\xi(t)$ is a Gaussian white noise with variance $\sigma^2$ and zero mean. The linear part of the model selects, by linear filtering, a subset of the possible stimuli probed by $I(t)$. That subset is expressed as $n$ relevant features $\{\varepsilon_\mu(t)\}, \quad (\mu = 1, 2, \ldots, n)$. Interpreted as vectors, the components of any stimulus that are relevant to changing the firing rate can be expressed in terms of projections onto these features. The firing rate of the model for a given temporal sequence $I(t)$ depends only on $\mathbf{s}$, the input filtered by the $n$ relevant features. Thus the firing rate from the given stimulus depends on the convolution of the input with all $n$ features and can be written as $P(\text{spike}\,|\,\mathbf{s} = I_0\bar{\boldsymbol{\varepsilon}} + \mathbf{x})$ where

$$\bar{\varepsilon}_\mu = \int_0^\infty d\tau\, \varepsilon_\mu(\tau), \quad x_\mu = \int_0^\infty d\tau\, \varepsilon_\mu(\tau)\xi(t-\tau).$$

Since $I(t)$ is white noise with stationary statistics, the projections $x_\mu$ can be taken to be stationary random variables chosen from a Gaussian distribution at each $t$.

Given the filtered stimulus, a nonlinear decision function $P(\text{spike}\,|\,I_0\bar{\boldsymbol{\varepsilon}} + \mathbf{x})$ generates the instantaneous time-varying firing rate. For a specified model and stimulus statistics, the mean firing rate $f(I_0, \sigma^2) = P(\text{spike})$ is simply

$$f(I_0, \sigma^2) = \int d\mathbf{s}\, P(\text{spike}\,|\,\mathbf{s})P(\mathbf{s}) = \int d\mathbf{x}\, P(\text{spike}\,|\,I_0\bar{\boldsymbol{\varepsilon}} + \mathbf{x})p(\mathbf{x}), \tag{9}$$

where

$$p(\mathbf{x}) = \frac{1}{(2\pi\sigma^2)^{n/2}} \cdot \exp\left(-\frac{1}{2\sigma^2}\|\mathbf{x}\|^2\right).$$

Eq. (9) describes an *f-I* curve of the model in the presence of added noise with variance $\sigma^2$. The slope or *gain* of the firing rate with respect to mean or variance can be computed if $P(\text{spike}\,|\,I_0\bar{\boldsymbol{\varepsilon}} + \mathbf{x})$ is known. However, the gains can be also obtained in terms of the first and second moments of $P(\text{spike}\,|\,I_0\bar{\boldsymbol{\varepsilon}} + \mathbf{x})$, which can be measured *directly* by reverse correlation analysis.

**Reverse correlation analysis**

We used spike-triggered reverse correlation to probe the computation of the model neurons through an LN model. We collected about 100,000 spikes and corresponding ensembles of spike triggered stimulus histories in a 30ms long time window preceding each spike.

From the spike-triggered input ensembles, we calculated spike-triggered averages (STA) and spike-triggered covariances (STC). The STA is simply the average of the set of stimuli that led to spikes subtracted from the mean of the "prior" stimulus distribution, the distribution of all stimuli independent of spiking output:

$$\mathrm{STA}(t) = \left\langle I(t_{\mathrm{spike}} - t) \right\rangle_{\mathrm{spike}} - \left\langle I \right\rangle_{\mathrm{prior}} = \left\langle \xi(t_{\mathrm{spike}} - t) \right\rangle_{\mathrm{spike}}, \tag{10}$$

Therefore, one may consider only the noise part of the zero mean stimulus.

When computing the STC, the prior's covariance is subtracted:

$$\begin{aligned} \Delta C(t,t') &= C_{\mathrm{spike}} - C_{\mathrm{prior}} \\ &= \left\langle \left\{ \xi(t_{\mathrm{spike}} - t) - \mathrm{STA}(t) \right\} \left\{ \xi(t_{\mathrm{spike}} - t') - \mathrm{STA}(t') \right\} \right\rangle_{\mathrm{spike}} - C_{\mathrm{prior}}. \end{aligned} \tag{11}$$

**Statistical analysis**

In calculating the slope and curvature of the *f-I* curves, we used 6–10 degree polynomial fitting of the *f-I* curves, where in any single case, the lowest degree was used which provided both a good fit and smoothness. From the fitting procedure, we obtained the standard deviation of the residuals, $\sigma_{\mathrm{fit}}$. This was repeated five times for *f-I* curves computed using different noise samples, and from this we computed $\sigma_{\mathrm{repeat}}$, the standard deviation of each computed slope and curvature. We estimated the total error of our calculation as $\sigma_{\mathrm{total}} = \sqrt{\sigma_{\mathrm{repeat}}{}^2 + \sigma_{\mathrm{fit}}{}^2}$. In practice, $\sigma_{\mathrm{repeat}}$ was always greater than $\sigma_{\mathrm{fit}}$ by an order of magnitude. This $\sigma_{\mathrm{total}}$ was used for the error bars in Fig. 3.

To evaluate the goodness of fit in Fig. 3, we used the Pearson $\chi^2$ test by using the reduced $\chi^2$ statistic

$$\chi^2 = \sum \frac{(O-E)^2}{\sigma_{\mathrm{total}}{}^2},$$

where $O$ and $E$ represent the right and left hand sides of Eqs. (4)–(6)respectively. From this, the $p-$values are estimated from the cumulative density function of the $\chi^2$ distribution, $Q(\chi^2 / k, k)$. The degree of freedom is $k = 54$ and $k = 34$ for the HH and HHLS, respectively.

**Derivation of Eqs. (4)-(6)**

We first present two key identities: the first one, which depends on the form of $\mathbf{s}$ having additive mean and noise components, is a change of variables for the gradient of $P(\text{spike} \mid \mathbf{x} + I_0\bar{\boldsymbol{\varepsilon}})$,

$$\frac{\partial P(\text{spike} \mid \mathbf{x} + I_0\bar{\boldsymbol{\varepsilon}})}{\partial I_0} = \sum_\mu \bar{\varepsilon}_\mu \frac{\partial P(\text{spike} \mid \mathbf{x} + I_0\bar{\boldsymbol{\varepsilon}})}{\partial x_\mu}. \tag{12}$$

Secondly, when $x$ is a Gaussian random variable with zero mean and variance $\sigma^2$, by using integration by parts in can be seen that any function $F(x)$ satisfies

$$\langle F'(x) \rangle = \frac{1}{\sigma^2}\langle xF(x) \rangle, \tag{13}$$

$$\begin{aligned}\langle F''(x) \rangle &= \frac{1}{\sigma^2}\langle [xF(x)]' \rangle - \frac{1}{\sigma^2}\langle F(x) \rangle \\ &= \frac{1}{\sigma^4}\langle x^2 F(x) \rangle - \frac{1}{\sigma^2}\langle F(x) \rangle.\end{aligned}$$

Then, we first take derivatives of both sides of Eq. (9) (or equivalently Eq. (1)), by $I_0$ and $\sigma^2$, and apply Eqs. (12)and(13). The first order in $I_0$ is

$$\begin{aligned}\frac{\partial \log f}{\partial I_0} = \frac{1}{f}\frac{\partial f}{\partial I_0} &= \frac{1}{f}\sum_\mu \bar{\varepsilon}_\mu \left\langle \frac{\partial}{\partial x_\mu} P(\text{spike} \mid \mathbf{x} + I_0\bar{\boldsymbol{\varepsilon}}) \right\rangle_{\mathbf{x}} \\ &= \frac{1}{\sigma^2}\cdot\frac{1}{f}\sum_\mu \bar{\varepsilon}_\mu \langle x_\mu P(\text{spike} \mid \mathbf{x} + I_0\bar{\boldsymbol{\varepsilon}}) \rangle_{\mathbf{x}}\end{aligned} \tag{14}$$

The second order is given by

$$\begin{aligned}\frac{\partial^2 \log f}{\partial {I_0}^2} &= \frac{1}{f}\frac{\partial^2 f}{\partial {I_0}^2} - \frac{1}{f^2}\left(\frac{\partial f}{\partial I_0}\right)^2, \\ \frac{\partial^2 f}{\partial {I_0}^2} &= \sum_{\mu,\nu} \bar{\varepsilon}_\mu \bar{\varepsilon}_\nu \left\langle \frac{\partial}{\partial x_\mu}\frac{\partial}{\partial x_\nu} P(\text{spike} \mid \mathbf{x} + I_0\bar{\boldsymbol{\varepsilon}}) \right\rangle_{\mathbf{x}} \\ &= \frac{1}{\sigma^4}\sum_{\mu,\nu} \bar{\varepsilon}_\mu \bar{\varepsilon}_\nu \langle (x_\mu x_\nu - \sigma^2 \delta_{\mu\nu}) P(\text{spike} \mid \mathbf{x} + I_0\bar{\boldsymbol{\varepsilon}}) \rangle_{\mathbf{x}},\end{aligned} \tag{15}$$

where $\delta_{\mu\nu}$ is a Kronecker delta symbol. The gain with respect to variance is

$$\frac{\partial f}{\partial \sigma^2} = -\frac{n}{2\sigma^2} f + \frac{1}{2\sigma^4} \sum_\mu \left\langle x_\mu^2 P(\mathrm{spike} \mid \mathbf{x} + I_0 \bar{\boldsymbol{\varepsilon}}) \right\rangle_x$$
$$= \frac{1}{2\sigma^4} \sum_\mu \left\langle \left( x_\mu^2 - \sigma^2 \right) P(\mathrm{spike} \mid \mathbf{x} + I_0 \bar{\boldsymbol{\varepsilon}}) \right\rangle_{\mathbf{x}}. \tag{16}$$

Now, we show how the right hand sides of Eqs. (14)-(16)correspond to the STA and the STC. Given a Gaussian white noise signal $\xi(t)$, we can split it as $\xi = \xi_{\parallel} + \xi_{\perp}$, where $\xi_{\parallel}$ belongs to the space spanned by our basis features $\{\varepsilon_\mu\}$, and therefore relevant to spiking. $\xi_{\perp}$ is the orthogonal or irrelevant part. $\xi_{\parallel}$ can be written as

$$\xi_{\parallel}(t) = \boldsymbol{\varepsilon} \cdot \mathbf{x} = \sum_\mu x_\mu \varepsilon_\mu, \quad x_\mu = \int_0^\infty d\tau \, \varepsilon_\mu(\tau) \xi(t - \tau).$$

Again, $\mathbf{x}$ is a Gaussian variable from a distribution Eq. (9).

The STA is

$$\mathrm{STA} = \langle \xi \rangle_{\mathrm{spike}} = \langle \xi_{\parallel} \rangle_{\mathrm{spike}} = \int d^n \mathbf{x} \, (\boldsymbol{\varepsilon} \cdot \mathbf{x}) P(\mathbf{x} + I_0 \bar{\boldsymbol{\varepsilon}} \mid \mathrm{spike}),$$

since $\xi_{\perp}$ is irrelevant and does not make any contribution. Here we use Bayes theorem,

$$\frac{P(\mathrm{spike} \mid \mathbf{x} + I_0 \bar{\boldsymbol{\varepsilon}})}{P(\mathrm{spike})} = \frac{P(\mathbf{x} + I_0 \bar{\boldsymbol{\varepsilon}} \mid \mathrm{spike})}{P(\mathbf{x} + I_0 \bar{\boldsymbol{\varepsilon}})},$$

As in Eq. (9), $P(\mathbf{s} = \mathbf{x} + I_0 \bar{\boldsymbol{\varepsilon}}) = p(\mathbf{x})$, and therefore the STA becomes

$$\mathrm{STA} = \int d^n \mathbf{x} \, (\boldsymbol{\varepsilon} \cdot \mathbf{x}) \frac{P(\mathrm{spike} \mid \mathbf{x} + I_0 \bar{\boldsymbol{\varepsilon}})}{P(\mathrm{spike})} p(\mathbf{x})$$
$$= \frac{1}{f} \sum_\mu \varepsilon_\mu \left\langle x_\mu P(\mathrm{spike} \mid \mathbf{x} + I_0 \bar{\boldsymbol{\varepsilon}}) \right\rangle_{\mathbf{x}}.$$

Comparing this result with Eq. (14), we obtain Eq. (4).

A similar calculation for the second order [19] shows

$$\Delta C(t, t') = \frac{1}{f} \sum_{\mu,\nu} \varepsilon_\mu(t) \varepsilon_\nu(t') \left\langle \left( x_\mu x_\nu - \sigma^2 \delta_{\mu\nu} \right) P(\mathrm{spike} \mid \mathbf{x} + I_0 \bar{\boldsymbol{\varepsilon}}) \right\rangle_{\mathbf{x}}$$
$$- \mathrm{STA}(t) \cdot \mathrm{STA}(t').$$

This result, combined with Eqs. (15)and (16), leads to (5) and (6), respectively.

**Acknowledgments**

This work was supported by a Burroughs-Wellcome Careers at the Scientific Interface grant, a Sloan Research Fellowship and a McKnight Scholar Award to ALF. BNL was supported by grant number F30NS055650 from the National Institute of Neurological Disorders and Stroke, the Medical Scientist Training Program at UW supported by the National Institute of General Sciences, and an ARCS fellowship.

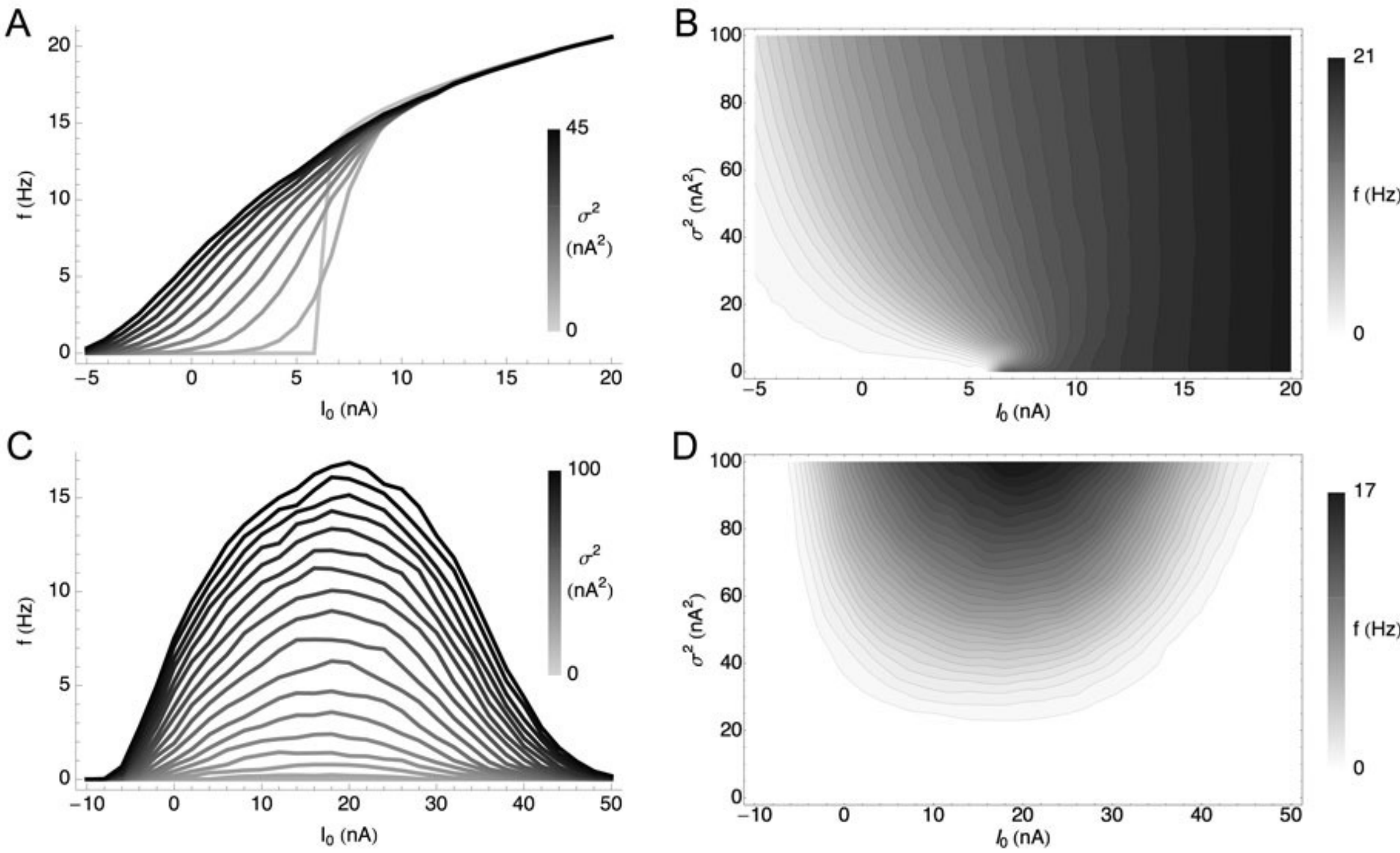

**Figure 1. Variance-dependent gain modulation of the HH and HHLS model.** Each model is simulated as described in Materials and Methods. **A.** *f-I* curves of a standard HH model for differing 10 variances ($\sigma^2$) from $0nA^2$ to $45nA^2$. The topmost trace is the response to the highest variance. Each curve is obtained with 31 mean values ($I_0$) ranging from -5nA to 20nA. **B.** The same data as A plotted in the (mean, variance) plane. Lighter shades represent higher firing rates. We used cubic spline interpolation for points not included in the simulated data. **C, D.** *f-I* curves of the HHLS model as in **A** and **B**. 10 means from -10nA to 50nA and 21 variances from $0nA^2$ to $100nA^2$ are used.

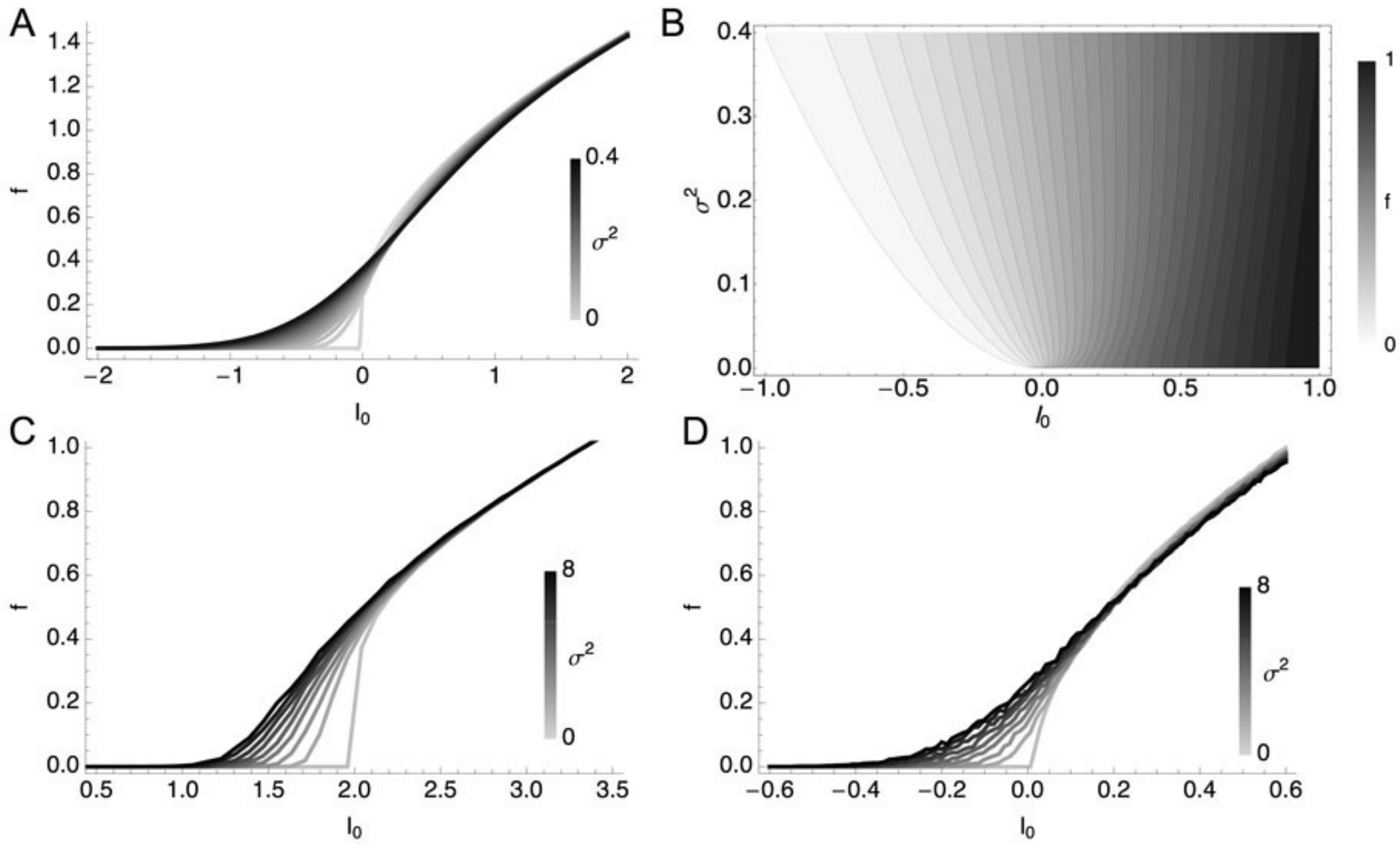

**Figure 2. Variance-dependent gain modulation of one-dimensional models. A.** Variance-dependent *f-I* curves of a one-dimensional model from the solution of Eq. (3) with the boundary condition, $f = \sqrt{I + 0.1}$ for $I > 0$ and $f = 0$ for $I \leq 0$ at zero noise.**B.** The firing rates of A in the (mean, variance) plane. **C.** *f-I* curves of an LIF model. **D.** *f-I* curves of a QIF model. The model parameters for the LIF and QIF are in Materials and Methods. We used 50 mean ($I_0$) values from 0 to 4 (LIF) and from -2 to 2 (QIF), and 8 variances ($\sigma^2$) from 0 to 8 for both models.

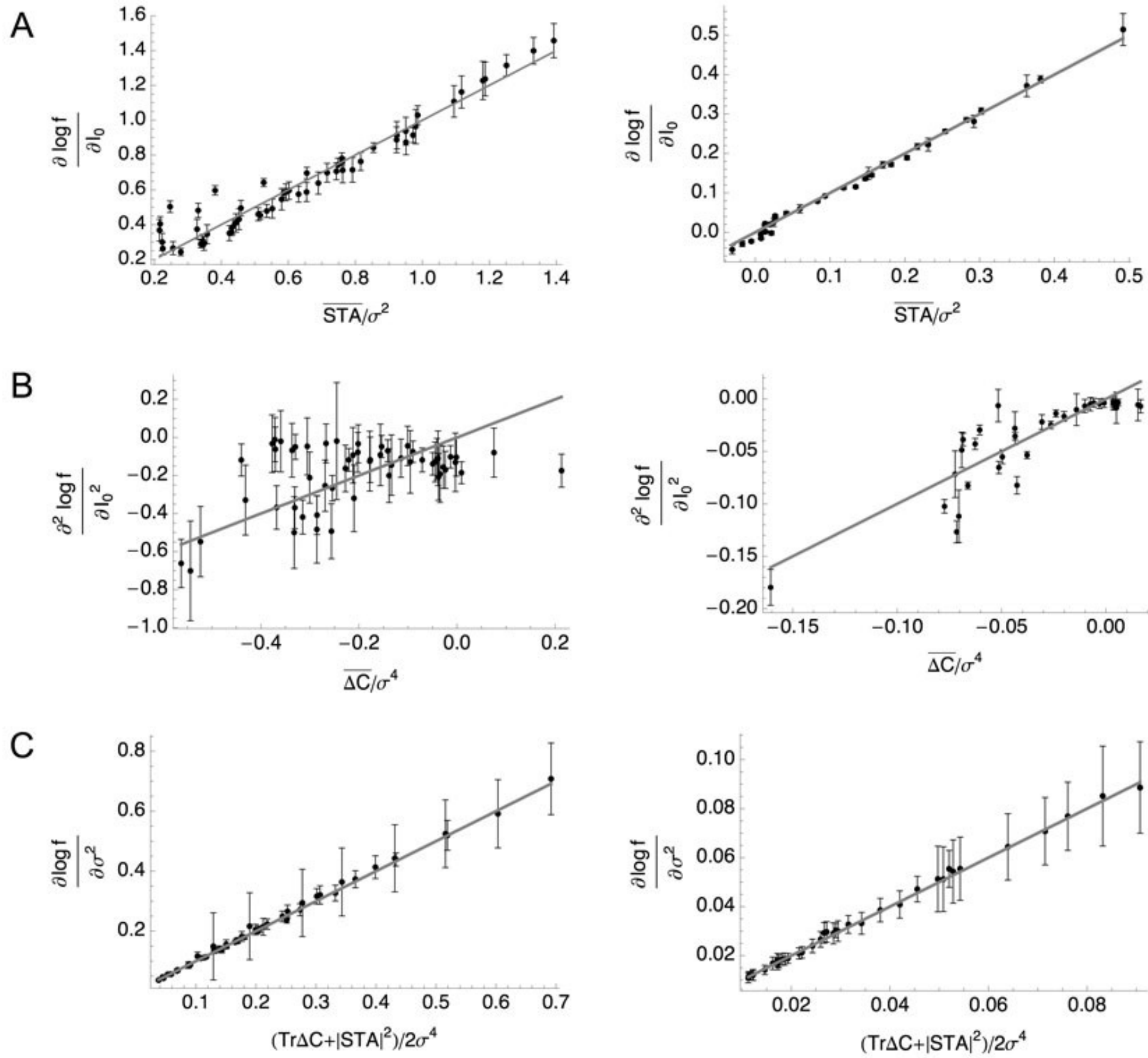

**Figure 3. Derivatives of the firing rate curves with respect to mean and variance related to quantities obtained by white noise analysis for the standard HH (left) and HHLS (right) models.** Each point is calculated from the simulated data with a selected (mean, variance) input parameter pair, as described in Materials and Methods, and the gray lines represent our theoretical predictions, Eq. (4)-(6), which hold when the variance dependent change in *f-I* curves is only due to intrinsic adaptation. A. Gain vs the norm of the STA, as in Eq. (4). **B.** Gain change vs the spike-triggered covariance term of Eq. (5). **C.** Change of firing rate with respect to variance vs. the function of the STA and spike-triggered covariance given in Eq. (6).

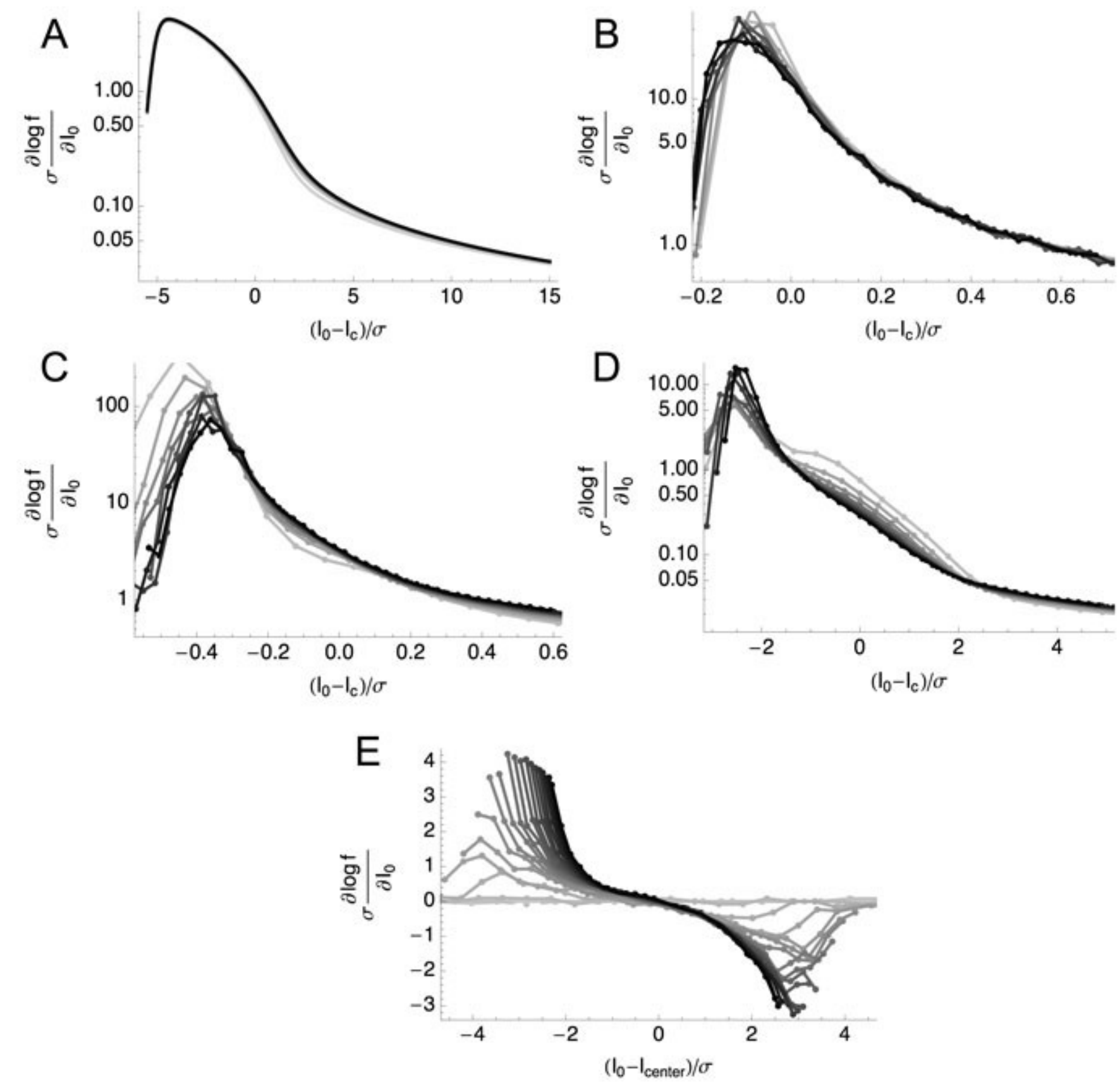

**Figure 4. Rescaled relative gains of variance-dependent *f-I* curves. A.** The one-dimensional LN ,**B.** QIF , and **C.** LIF model. The same data as Fig. 2 are used. **D.** The standard HH model from Fig. 1A and B. **E.** The HHLS model from Fig. 1C and D. Since the HHLS does not have a rheobase, we instead used $I_{\mathrm{center}}$ = 20nA at which the variance-dependent firing rate increase is maximal.

# Supporting Information

*Intrinsic gain modulation and adaptive neural coding*

Sungho Hong, Brian Nils Lundstrom and Adrienne L. Fairhall

## Firing rate of the LIF model with noisy stimuli

There are known analytic formulas for the firing rate of the simple neuron models given Gaussian noise stimuli such as the LIF [1-3] and QIF [4,5]. Fourcaud-Trocmé et al. [6] obtained a formula for a large class of models including the LIF and QIF. Here we use them to discuss two aspects of the LIF model.

First, we show that the firing rate always increases as variance increases. The analytic form of the firing rate is [1-3]

$$f(I_0,\sigma^2)=\left(\tau\int_{\frac{CV_r-I_0\tau}{\sigma\tau}}^{\frac{C\theta-I_0\tau}{\sigma\tau}}\sqrt{\pi}e^{x^2}\left(1+\mathrm{erf}(x)\right)dx\right)^{-1}, \tag{17}$$

where $\tau=1/g_L$. For convenience, we define $J(x)\equiv e^{x^2}\left(1+\mathrm{erf}(x)\right)$. The firing rate change with variance is given by

$$\begin{aligned}\frac{1}{f^2}\frac{\partial f}{\partial\sigma^2}&=\frac{\tau\sqrt{\pi}}{2}\left(\frac{C\theta-I_0\tau}{\sigma^3\tau}\right)\cdot J\left(\frac{C\theta-I_0\tau}{\sigma\tau}\right)-\frac{\tau\sqrt{\pi}}{2}\left(\frac{CV_r-I_0\tau}{\sigma^3\tau}\right)\cdot J\left(\frac{CV_r-I_0\tau}{\sigma\tau}\right)\\&=\frac{\tau\sqrt{\pi}}{2\sigma^2}\,xJ(x)\Bigg|_{\frac{CV_r-I_0\tau}{\sigma^3\tau}}^{\frac{C\theta-I_0\tau}{\sigma\tau}}.\end{aligned}$$

Now $xJ(x)$ is an increasing function of $x$ whose minimum is $\lim_{x\to-\infty}xJ(x)=-\frac{1}{\sqrt{\pi}}$. Therefore, $\frac{\partial f}{\partial\sigma^2}$ is always positive and the firing rate also always increases with variance.

However, with a larger variance, the change in the firing rate becomes smaller and the firing rate approaches an asymptotic limit. In the limit $\sigma^2\to\infty$ and $I_0/\sigma\to\text{finite}$, Eq. (17) vanishes in the leading order. The next leading order survives as

$$f = \left[ C\left( \frac{\theta - V_r}{\sigma} \right) J\left(-I_0 / \sigma\right) \right]^{-1}. \tag{18}$$

Note that Eq. (18) is factorized in a similar way to Eq. (8) in the paper. Therefore, the rescaled relative gain is

$$\frac{\sigma}{f}\frac{\partial f}{\partial I_0} = \frac{J'(-I_0 / \sigma)}{J(-I_0 / \sigma)},$$

which is a function only of $I_0 / \sigma$ as we have seen in Fig. 3C. Note that the firing rate in this limit has a form of a function of $I_0 / \sigma$ multiplied by a factor which only depends on the variance, and this makes the rescaled relative gain only a function of $I_0 / \sigma$.

In the QIF case, the firing rate is [4,5]

$$f = \frac{\sigma^{3/2}}{\pi^{1/2}} \left( \int_{-\infty}^{\infty} d\zeta \exp\left[ -\frac{1}{\sigma^{1/2}} \left( (I_0 / \sigma)^2 \zeta^2 + \zeta^6 / 48 \right) \right] \right)^{-1}$$

where we have taken $C = 1$ and $L(V) = V^2$ for simplicity. Here, we do not have a simple factorization as Eq. (18) in the $\sigma \to \infty$ limit. Therefore, the rescaling is not directly related to its dynamics, but is rather phenomenological and approximate.